\documentclass[12pt]{article}
\usepackage{booktabs}
\usepackage{latexsym}
\usepackage{tikz}
\usetikzlibrary{arrows,automata}
\usepackage{graphicx}
\usepackage{dcolumn}

\usepackage{amssymb}
\usepackage{amsmath}

\usepackage{tikz}
\usepackage{tikz,tikz-3dplot} 
\usetikzlibrary{quantikz}
\usepackage{tikz-qtree}
\usetikzlibrary{trees}

\usepackage{hyperref}
 
\usepackage{pgfplots}
 \pgfplotsset{compat = newest}

 \usepackage{xcolor}
\usepackage{colortbl}
\newcolumntype{a}{>{\columncolor{blue}}c}

\usetikzlibrary{calc}

\begin{document}

\def\myvdots{\ \vdots\ }

\pagestyle{plain}

\pagenumbering{arabic}

\setlength{\parindent}{0in}

\setlength{\parskip}{2.2ex}

\setlength{\footskip}{.6in}
\begin{footnotesize}

\begin{center}
 {\bf Twisted Fiber Bundle Codes over Group Algebras}\\
{Chaobin Liu}\footnote{Department of Mathematics, Bowie State University, MD, USA\\ Email: cbliu2000@yahoo.com}
\end{center}

\begin{center}
 Abstract
\end{center}

We introduce a twisted fiber bundle construction of quantum CSS codes over group algebras \(R=\mathbb F_2[G]\), where each base generator carries a generator-dependent \(R\)-linear fiber twist satisfying a flatness condition. This construction extends the untwisted lifted product code, recovered when all twists are identities. We show that invertible twists (satisfying a flatness condition) give a complex chain-isomorphic to the untwisted one, so the resulting binary CSS codes have the same blocklength \(n\) and encoded dimension \(k\). In contrast, singular chain-compatible twists can lower boundary ranks and increase the number of logical qubits. Examples over \(R=\mathbb F_2[D_3]\) show that singular chain-compatible twists can increase the encoded dimension \(k\) at fixed blocklength \(n\), and in these finite examples the minimum distance \(d\) remains unchanged. This provides evidence that singular twisting enlarges the design space beyond the ordinary lifted product construction.

\section{Introduction}


Homological and product-type constructions have become a major source of quantum CSS codes \cite{NC2000}. In particular, hypergraph product \cite{TZ2013}, lifted product \cite{PK2019,PK2021}, balanced product \cite{BE2021}, and fiber bundle-inspired constructions \cite{HHD2021} provide systematic frameworks for constructing low-density parity-check (LDPC) CSS codes from algebraic data. These approaches have produced many important families of codes, including those in \cite{BEPRX2021, AF2024, BH2014, MC2025, PK2022, EKZ2020, KT2020, FGL2018, ZP2019, LAV2020, LZ2022, DHLV2023} and the references therein. In many cases, they proceed by combining two smaller chain complexes into a larger complex whose first homology determines the logical qubits. Their appeal lies in the fact that questions about code parameters can then be reformulated in terms of the algebraic and homological properties of the underlying complexes \cite{CE1999, W1994}.

In the present work, we study a twisted fiber bundle variant of this philosophy over group algebras.  Let $R=\mathbb F_2[G]$
for a finite group \(G\).  Starting from two 2-term chain complexes of free modules over \(R\),
\(
B:\; B_1 \xrightarrow{\partial^B} B_0\) and \(F:\; F_1 \xrightarrow{\partial^F} F_0,\)
we construct a \(3\)-term complex
\[
C_2 \xrightarrow{\partial_2} C_1 \xrightarrow{\partial_1} C_0, \,\,
C_2=B_1\otimes_R F_1,\,\,
C_1=(B_1\otimes_R F_0)\oplus(B_0\otimes_R F_1),\,\,
C_0=B_0\otimes_R F_0,
\]
in which the interaction between the base $B$ and fiber $F$ is governed by generator-dependent \(R\)-linear twist maps.  Concretely, each basis generator of \(B_1\) is equipped with a pair of fiber maps
\(
\varphi_{0,j}:F_0\to F_0\) and 
\(\varphi_{1,j}:F_1\to F_1,
\)
and the total differentials are defined by inserting these twists into the base component of the boundary operator. The resulting construction may be viewed as a twisted fiber bundle code \cite{HHD2021} at the module level, with the ordinary lifted product code \cite{PK2019, PK2021} appearing as the special case in which all twists are identities.

The first basic requirement is that the twisted total differentials still satisfy the chain condition.  In our setting, where the coefficients of \(\partial^B\) act by the right regular action and the fiber differential \(\partial^F\) together with the twists \(\varphi_{0/1,j}\) act by the left regular action, the condition \(\partial_1\partial_2=0\) reduces to a generatorwise flatness relation \footnote{The flatness relation/condition requires that the degree-1 and degree-0 twist maps (i.e., \(\varphi_{1,j}\) and \(\varphi_{0,j}\)) commute with the fiber differential, so that each twist acts as a chain map and the resulting twisted total complex satisfies 
\(\partial^2=0\).}
\[
\varphi_{0,j}\partial^F=\partial^F\varphi_{1,j}
\,\,\text{for all }j.\]

Thus the twisting data may be interpreted as a flat connection on the fiber complex, indexed by the generators of the base complex.  This viewpoint makes clear that the lifted product construction corresponds to the trivial flat connection, while the twisted construction allows the fiber transport to vary from generator to generator.

A natural question is whether such twisting produces genuinely new quantum codes, or merely reparameterizes existing lifted product constructions.  When the twists are invertible, one may ask whether they can be removed by a fiberwise change of basis.  In that case the twisted complex is chain-isomorphic to the untwisted one.  Consequently, the corresponding binary CSS codes have the same blocklength \(n\) and encoded dimension \(k\), and in many cases are equivalent under a coordinate transformation.  This shows that invertibility alone may not be enough to produce new parameters: the key issue is whether the twist data is globally removable or genuinely nontrivial.

The more interesting phenomenon arises when one relaxes invertibility and allows the twists to be merely chain-compatible endomorphisms.  In that regime, the twist maps may develop kernel and cokernel defect spaces, and these defects can alter the homology of the total complex.  At an algebraic level, singular twists weaken the effective action of the induced boundary operators and may lower their ranks.  From the coding-theoretic point of view, this can enlarge the logical subspace by creating additional nontrivial homology classes.  Put differently, singular twists act as a source of homological amplification: they may create new cycles and simultaneously prevent some of them from becoming boundaries.


Our concrete examples are carried out over the nonabelian group algebra \(R=\mathbb F_2[D_3]\),
where the small size of \(D_3\) (symmetry group of equilateral triangle) makes explicit computation feasible while still retaining genuinely noncommutative behavior. These examples illustrate several distinct regimes. In some cases, invertible twists produce no change in the binary code parameters, reflecting that the twists may be just a different presentation of the same complex. In others, noninvertible but chain-compatible twists lower the ranks of the boundary maps and increase the encoded dimension \(k\).  Most notably, we exhibit finite-length examples in which the twisted code has strictly larger \(k\) than the corresponding untwisted lifted product code while preserving the same blocklength \(n\) and, in our examples, the same minimum distance \(d\).  Thus, at least at the finite-length level, singular twisting can improve the rate without sacrificing distance.

These examples also show that twisted fiber bundle codes with generator-dependent \(R\)-module twists need not be equivalent to lifted product codes.  Indeed, once singular twists are admitted, the resulting homological structure can differ essentially from that of the untwisted complex.  This enlarges the design space beyond the usual lifted product setting and suggests that noninvertible twisting may be a useful mechanism for engineering additional logical qubits in algebraic CSS constructions.

The main contributions of this paper are as follows.  First, we formulate a twisted fiber bundle construction over arbitrary group algebras \(R=\mathbb F_2[G]\), together with the corresponding flatness condition ensuring that the total complex is well defined.  Then, we identify the invertible regime in which the twisted and untwisted constructions are chain-isomorphic. Finally, we support the general framework with explicit computations over \(\mathbb F_2[D_3]\), showing that singular twisting can strictly improve \(k\) relative to the corresponding lifted product code at fixed \(n\) and, in the examples presented here, fixed \(d\).

The paper is organized as follows. Section~\ref{prel} reviews the necessary background on group algebras, chain complexes, and CSS codes. Section~\ref{construction} introduces the twisted fiber bundle construction and derives the associated chain-compatibility condition. Section~\ref{isomorphism} analyzes the case of invertible twists and explains its relation to untwisted lifted product codes. Section~\ref{example} presents explicit examples over \(\mathbb F_2[D_3]\) and compares the parameters of the resulting twisted and untwisted codes. Detailed computations of the code parameters \([[n,k,d]]\) for both the untwisted code and the code with all singular twists are given in Appendix~C. Section~\ref{conclusion} concludes with remarks on the twisted fiber bundle code construction and an open problem. Finally, Appendix~D reviews lifted product codes, while Appendix~E reviews fiber bundle codes over the field \(\mathbb F_2\).


\section{Preliminaries} \label{prel}

{\bf Group algebras}: Let \(G\) be a finite group and let \(\mathbb{F}_2\) denote the field with two elements. {\it The group algebra} \(\mathbb{F}_2[G]\) is the set of all formal sums
\(\sum_{g\in G} a_g g\) where \(a_g\in \mathbb{F}_2.\)
Addition is defined coefficientwise, and multiplication is extended linearly from the multiplication in \(G\). Thus \(\mathbb{F}_2[G]\) is a finite-dimensional \(\mathbb{F}_2\)-algebra of dimension \(\ell=|G|\).

Elements of \(\mathbb{F}_2[G]\) may be viewed as algebraic combinations of group elements, and modules over \(\mathbb{F}_2[G]\) encode linear actions of \(G\). In particular, free \(\mathbb{F}_2[G]\)-modules provide a convenient framework for describing structured matrices whose entries carry group symmetry. After choosing a basis and applying a regular representation of \(G\), matrices over \(\mathbb{F}_2[G]\) can be converted into binary matrices over \(\mathbb{F}_2\), which is useful for constructing binary quantum codes.

{\bf Chain complexes}:\,\, A {\it chain complex} \(C_\bullet\)  is a sequence of modules (or vector spaces) and homomorphisms
\(\cdots \xrightarrow{\partial_{i+1}} C_i \xrightarrow{\partial_i} C_{i-1} \xrightarrow{\partial_{i-1}} \cdots\)
such that \(\partial_i \partial_{i+1} = 0\) for all \(i\).
The maps \(\partial_i\) are called {\it boundary operators} or {\it differentials}.

The condition \(\partial_i \partial_{i+1}=0\) implies that every boundary is a cycle. This leads to the \(i\)-th homology group
\(H_i(C_\bullet)=\ker(\partial_i)\big/\operatorname{im}(\partial_{i+1}),\) which measures the nontrivial cycles that are not boundaries.

In CSS code constructions, one often considers a \(3\)-term complex
\(C_2 \xrightarrow{\partial_2} C_1 \xrightarrow{\partial_1} C_0\) with \(\partial_1\partial_2=0.\)
When the modules are free over \(\mathbb{F}_2\) or over a group algebra \(\mathbb{F}_2[G]\), the boundary maps can be represented by matrices. The middle homology \(H_1(C_\bullet)\) then plays a central role in determining the number of encoded qubits.

Let \(D_\bullet: \cdots \xrightarrow{\partial^D_{i+1}} D_i \xrightarrow{\partial^D_i} D_{i-1} \xrightarrow{\partial^D_{i-1}} \cdots\) be another chain complex.

A {\it chain isomorphism} \(f_\bullet : C_\bullet \to D_\bullet\) is a family of module (or vector space) isomorphisms
\(f_i : C_i \to D_i\) where \(i\in \mathbb Z\) such that for every \(i\), \(\partial_i^D \circ f_i = f_{i-1} \circ \partial_i^C.\)

If two chain complexes are chain-isomorphic, then they have exactly the same homological information \cite{CE1999, W1994}. In particular, \(H_n(C_\bullet)\cong H_n(D_\bullet)\) for all \(n\). Thus, two chain complexes are chain-isomorphic if they differ only by degreewise invertible changes of coordinates compatible with the differentials.

{\bf CSS codes}: A \emph{CSS quantum code} is specified by two binary parity-check matrices
\(H_X\) and \( H_Z\) satisfying the orthogonality condition \(H_X H_Z^T = 0.\) This condition ensures that the corresponding \(X\)-type and \(Z\)-type stabilizer generators commute.

If the code has \(n\) physical qubits, then \(n\) is the number of columns of \(H_X\) and \(H_Z\). The number of encoded qubits is
\(k = n - \operatorname{rank}(H_X) - \operatorname{rank}(H_Z).\) The distance is determined by the minimum weight of a nontrivial logical \(X\)- or \(Z\)-operator.

A \(3\)-term chain complex of vector spaces over \(\mathbb{F}_2\) naturally gives rise to a CSS code by setting
\(H_X = \partial_1\) and \(H_Z = \partial_2^\top.\)
Then the chain condition \(\partial_1\partial_2=0\) is exactly the CSS commutation condition
\(H_X H_Z^T = 0.\) In this correspondence, the code length is
\(n = \dim C_1,\) and the number of encoded qubits is \(k = \dim H_1(C_\bullet).\)

Thus, chain complexes provide a natural homological framework for constructing CSS codes, while group algebras provide additional algebraic structure and symmetry that can be exploited in code design.

\section{Twisted fiber bundle code construction} \label{construction}


Let \(G\) be a finite group and let \(R=\mathbb F_2[G]\) denote its group algebra. All modules below are free left \(R\)-modules.
We construct a 3-term {\it quasi} chain complex
\(C_\bullet: C_2 \xrightarrow{\partial_2} C_1 \xrightarrow{\partial_1} C_0\)
of free \(R\)-modules \footnote{We note that, when an associative algebra \(R\) is noncommutative, this 3-term sequence of \(R\)-modules need not be a chain complex over \(R\). However, after applying the flatness condition and compatible left and right regular representations, its binary expansion is a genuine chain complex; see Appendix A for the proof. We therefore refer to the original sequence as a 3-term quasi chain complex over \(R\). Throughout this work, unless otherwise stated, any 3-term chain complex over \(R\) is to be understood in this quasi chain complex sense.} defining a CSS code.

1. Base and Fiber Complexes

Base complex:\quad Let \(B: B_1 \xrightarrow{\partial^B} B_0\) be a 2-term chain complex of free \(R\)-modules.
We choose free bases \(B_1 \cong R^{m}\) and \(B_0 \cong R^{n}\), and write \(\partial^B \in \mathcal{M}_{n\times m}(R).\)

Fiber complex:\quad Let \(F:F_1 \xrightarrow{\partial^F} F_0\) be another 2-term chain complex of free \(R\)-modules with \(F_1 \cong R^{p}\) and \(F_0 \cong R^{q}\),
and \(\partial^F \in \mathcal{M}_{q\times p}(R).\)
We assume that \(F\) is equipped with the associative ring of \(R\)-linear chain endomorphisms\footnote{For an \(R\)-module \(M\), the set
\(\operatorname{End}_R(M)=\{\varphi:M\to M \mid \varphi \text{ is an } R\text{-module homomorphism}\}\)
is an associative ring under the usual addition of homomorphisms and multiplication given by composition.}
\[
\operatorname{End}_R(F)
=
\left\{
(\varphi_1,\varphi_0)
\;\middle|\;
\varphi_i\in \operatorname{End}_R(F_i),\ 
\varphi_0\partial^F=\partial^F\varphi_1
\right\}.
\]

Let \({e_1,\dots,e_m}\) be the chosen basis of \(B_1\). To each base generator \(e_j\), we assign a chain endomorphism
\(\varphi_j=(\varphi_{1,j},\varphi_{0,j})\in \mathrm{End}_R(F)\). Thus \(\varphi_{0,j}\partial^F= \partial^F \varphi_{1,j}\) for each\, \(j\).

3. Total Module Structure

We define free \(R\)-modules as follows:
\begin{eqnarray} \nonumber
\!\!C_2:
\!&=&B_1 \otimes_R F_1\\ \nonumber
\!\!C_1:
\!&=& (B_1 \otimes_R F_0)\oplus (B_0 \otimes_R F_1)\\ \nonumber
\!\!C_0:
\!&=&B_0 \otimes_R F_0 \nonumber
\end{eqnarray}
Since tensor products of free \(R\)-modules are free, all \(C_i\) are free \(R\)-modules with ranks:\, \(\operatorname{rank}_R C_2 = mp\), \(\operatorname{rank}_R C_1 = mq + np\),\, and \(\operatorname{rank}_R C_0 = nq.\)

4. Twisted Boundary Maps

The boundary \(\partial_2: C_2 \to C_1\) is defined by
\(\partial_2=\begin{bmatrix}\mathrm{id}_{B_1}\otimes \partial^F \\
\partial^B_{\varphi_1}\end{bmatrix}.\)
The new ingredient is the twisted base component
\[\partial^B_{\varphi_1}:B_1\otimes_R F_1\longrightarrow B_0\otimes_R F_1,\] which is defined on basis elements by
\(\partial^B_{\varphi_1}(e_j\otimes f^1)=\sum_{i=1}^n
(\partial^B)_{i j}e_i^{\prime}\otimes\varphi_{1,j}(f^1),\)
where \(\{e_i^{\prime}\}\) is the chosen basis of \(B_0\), and \(f^1\in F_1\).

Thus the fiber map applied depends on the column index \(j\), and different base generators twist fiber module \(F_1\) differently. In block-matrix form, \(\partial^B_{\varphi_1}=[(\partial^B)_{ij}\varphi_{1,j}]_{i,j}.\)

The boundary \(\partial_1: C_1 \to C_0\) is defined by
\(\partial_1=\begin{bmatrix}
\partial^B_{\varphi_0}&\mathrm{id}_{B_0}\otimes \partial^F\end{bmatrix}.\) The new ingredient is the twisted base component
\(\partial^B_{\varphi_0}:B_1\otimes_R F_0\longrightarrow B_0\otimes_R F_0,\) which is defined on basis elements by
\(\partial^B_{\varphi_0}(e_j\otimes f^0)=\sum_{i=1}^n
(\partial^B)_{i j}e_i^{\prime}\otimes\varphi_{0,j}(f^0),\)
where \(\{e_i^{\prime}\}\) is the chosen basis of \(B_0\), and \(f^0\in F_0\).

Similarly, the fiber map applied depends on the column index \(j\), and different base generators twist fiber module \(F_0\) differently. In block-matrix form, \(\partial^B_{\varphi_0}=[(\partial^B)_{ij}\varphi_{0,j}]_{i,j}.\)


5. The Resulting CSS Code

The resulting CSS code is defined from the binary boundary maps obtained after applying compatible left and right regular representations. Here, compatibility refers to the convention that coefficients arising from the base complex are represented by right regular representations, whereas coefficients arising from the fiber complex and the twisting maps are represented by left regular representations. With this convention and the flatness condition, the resulting binary matrices compose to zero.

To obtain the binary parity-check matrices, we proceed as follows.

First, we compute the twisted boundary matrices \(\partial_1\) and \(\partial_2\) over \(R\). Each \(R\)-entry appearing in these matrices is of the form \(b\omega\), \(f\), or \(0\), where \(b\), \(\omega\), and \(f\) are entries of \(\partial^B\), \(\varphi_{0/1,j}\), and \(\partial^F\), respectively.

Next, we apply the binary expansion. After fixing a basis of \(R\) over \(\mathbb F_2\), we replace each coefficient coming from the base complex by its right regular representation, and each coefficient coming from the fiber complex or the twisting maps by its left regular representation. In particular, a coefficient of the form \(b\omega\), arising from the interaction between a base boundary coefficient and a twisting coefficient, is represented by the product of the corresponding right and left regular matrices.
\[
b\omega \quad \longmapsto \quad \rho_b\lambda_\omega ,
\]
where \(\rho_b:x\mapsto xb\) is the right regular representation and \(\lambda_\omega:x\mapsto \omega x\) is the left regular representation. Since right and left regular representations commute, this convention is well defined at the binary matrix level.

Finally, we assemble the resulting block matrices to obtain the binary parity-check matrices \(H_X\) from \(\partial_1\) and \(H_Z\) from \(\partial_2^\top\). The CSS condition \(H_XH_Z^\top=0\) then follows from the flatness condition and the commutation of the left and right regular actions. A detailed verification is given in Appendix A.

The CSS code defined by the parity-check matrices \(H_X\) and \(H_Z\) is called a twisted fiber bundle code over the group algebra \(\mathbb F_2[G]\). Its length is \(n=\ell \operatorname{rank}_R C_1,\) where \(\ell=\dim_{\mathbb F_2}R=|G|\), and its dimension is \(k=n-\operatorname{rank}_{\mathbb F_2}H_X-\operatorname{rank}_{\mathbb F_2}H_Z.\)
The distance \(d\) depends on the interaction among the base homology, the fiber homology, and the twist endomorphisms \(\varphi_{0/1,j}\).

Remark. This construction generalizes the untwisted lifted product construction, which is recovered when all twists are trivial, i.e., when \(\varphi_{0/1,j}=\mathrm{id}\) for every base generator index \(j\) \cite{PK2019,PK2021}. Since the twisting maps may vary across base generators, one may ask whether the resulting code remains an LDPC CSS code. The answer is affirmative under the following conditions: each twist \(\varphi_{0/1,j}\) is sparse over \(R\), in the sense of having bounded row and column weight after binary expansion, and each base generator interacts with only a bounded number of fiber generators. In addition, the twists must satisfy the flatness condition in order to preserve CSS commutation. If the flatness condition fails, then the CSS condition may fail; if the sparsity condition fails, then the resulting CSS code may no longer be LDPC.

\section{Chain Isomorphism Between Twisted and Untwisted Lifted Product Chain Complexes}
\label{isomorphism}

When all twists are invertible, the twisted complex is chain-isomorphic to the corresponding untwisted complex. Consequently, the associated CSS codes have the same length and the same dimension. We formalize these observations in the following theorem and its corollaries.

\medskip

\noindent
{\bf Theorem}~(Chain isomorphism with mixed left/right regular actions).

Let \(R=\mathbb F_2[G]\) or, more generally, any associative ring for which the relevant regular actions are defined. Let \(B:B_1 \xrightarrow{\partial^B} B_0\) and \(F: F_1 \xrightarrow{\partial^F} F_0\)
be $2$-term chain complexes of free left $R$-modules.

Fix a basis $e_1,\dots,e_m$ of $B_1$ and a basis $e_1',\dots,e_n'$ of $B_0$. For each $j\in\{1,\dots,m\}$, let \(\varphi_{0,j}:F_0\to F_0\) and \(\varphi_{1,j}:F_1\to F_1\) be invertible left $R$-linear maps satisfying the flatness condition \(\varphi_{0,j}\partial^F=\partial^F\varphi_{1,j}.\)

Define the twisted product complex $C_\bullet^{\mathrm{tw}}$
\[
C_2=B_1\otimes_R F_1,
\qquad
C_1=(B_1\otimes_R F_0)\oplus(B_0\otimes_R F_1),
\qquad
C_0=B_0\otimes_R F_0,
\]
with differentials
\(
\partial_2^{\mathrm{tw}}=
\begin{bmatrix}
\mathrm{id}_{B_1}\otimes \partial^F\\[2mm]
\partial^B_{\varphi_1}
\end{bmatrix}\) and
\(\partial_1^{\mathrm{tw}}=
\begin{bmatrix}
\partial^B_{\varphi_0} & \mathrm{id}_{B_0}\otimes \partial^F
\end{bmatrix}
\) where
\(\partial^B_{\varphi_l}(e_j\otimes x)=\sum_{i=1}^n (\partial^B)_{ij}\, e_i' \otimes \varphi_{l,j}(x)\) for \(l=0, 1\).

Let $C_\bullet^{\mathrm{untw}}$ denote the corresponding untwisted lifted product complex with differentials
\(
\partial_2^{\mathrm{untw}}=
\begin{bmatrix}
\mathrm{id}_{B_1}\otimes \partial^F\\[2mm]
\partial^B\otimes \mathrm{id}_{F_1}
\end{bmatrix}\) and \(\partial_1^{\mathrm{untw}}=
\begin{bmatrix}
\partial^B\otimes \mathrm{id}_{F_0} & \mathrm{id}_{B_0}\otimes \partial^F
\end{bmatrix}.\)


Then $C_\bullet^{\mathrm{tw}}$ and $C_\bullet^{\mathrm{untw}}$ are chain-isomorphic. In particular, \(H_1(C^{\mathrm{tw}}_\bullet)\cong H_1(C^{\mathrm{untw}}_\bullet).\)

We defer the proof of this theorem to Appendix B.

{\bf Corollary}\,(equality of $n$ and $k$)
Under the hypotheses of the theorem, the twisted and untwisted lifted product binary CSS codes have the same length $n$ and the same dimension $k$.

{\bf Proof}\quad The two complexes have the same underlying module sizes, so the binary-expanded physical space has the same dimension in both cases, hence the same code length $n$. Under binary expansion, the right regular action used for the base coefficients $(\partial^B)_{ij}$ commutes with the left regular action used for $\partial^F$ and the twists $\varphi_{0/1,j}$. Hence the identities used in the proof of the theorem remain valid after binary expansion, so the resulting binary complexes are again chain-isomorphic, their first homology groups are isomorphic, so the number of encoded qubits (or the homological dimension) $k$ is the same in both cases.



{\bf Remark}\,(equality of distance is not automatic from chain-isomorphism)
The theorem proves equality of $n$ and $k$, but not automatically of the minimum distance $d$.

Indeed, the chain isomorphism on the middle space (see Appendix B for complete definitions) is
\(T_1=\mathrm{diag}(\varphi_{0,1}^{-1}, \varphi_{0,2}^{-1}, ..., \varphi_{0,m}^{-1})\oplus I\)
with respect to \(C_1=(B_1\otimes_R F_0)\oplus(B_0\otimes_R F_1).\)
After binary expansion, this map need not preserve Hamming weight on all vectors unless each $\varphi_{0,j}$ induces a binary monomial transformation (equivalently, permutation-type over \(\mathbb{F}_2\)) on its corresponding block. Thus equality of $d$ requires an additional argument.

{\bf Proposition}\,(sufficient condition for equality of $d$)
Assume, in addition to the hypotheses of the theorem, that each $\varphi_{0,j}$ induces a binary monomial map on the corresponding coordinate block of $B_1\otimes_R F_0$. Then the induced map $T_1$ on the binary-expanded middle space preserves Hamming weight. Therefore the associated twisted and untwisted CSS codes have the same minimum distance: \(d_{\mathrm{tw}}=d_{\mathrm{untw}}.\)

{\bf Proof}\quad A binary monomial map is a coordinate permutation; over $\mathbb F_2$, such a map preserves Hamming weight exactly. Since $T_1$ induces an isomorphism between the logical spaces and preserves the weight of every representative, it preserves the minimum weight of every nonzero logical class. Hence the code distances coincide.

{\bf Conjecture}\quad Invertibility and chain-compatibility may already imply equality of the full CSS parameters $[[n,k,d]]$. The theorem above proves this for $n$ and $k$, but the general distance statement remains open unless one proves that the induced middle-space isomorphism preserves minimum logical weight, or else finds a counterexample.

\section{Evidence that singular twists can improve the code rate} \label{example}

When all twists are invertible and chain-compatible, the two codes necessarily share the same parameters \(n\) and \(k\), as discussed above. A natural question then arises: can a twisted code outperform the corresponding untwisted lifted product code? We answer this by providing an example in which the twists are not all invertible but remain chain-compatible, resulting in an increase in \(k\) while \(d\) stays unchanged.



{\bf Example}:\,\,In the following example, where both the base and fiber are \(2\times 2\), we examine the code parameters according to three distinct cases determined by the twists' invertibility:

Case 1: both twists are invertible; Case 2: one twist is invertible and the other is not; Case 3: neither twist is invertible.


Let \(R=\mathbb F_2[D_3]\) where \(D_3=\langle r,s\mid r^3=s^2=1,\ srs=r^{-1}\rangle\). We choose two chain complexes \(B: B_1=R^2\xrightarrow{\partial^B}B_0=R^2\) and \(F:F_1=R^2\xrightarrow{\partial^F}F_0=R^2\) with differentials

\[
\partial^B=
\begin{bmatrix}
0 & r+r^2\\
1+r+r^2 & 0
\end{bmatrix}
\text{and}\,\, 
\partial^F=
\begin{bmatrix}
1&r\\
s&1
\end{bmatrix}.
\]
We check everything using the convention: base coefficients from \(\partial^B\) act by the right regular action, and fiber differential and twists act by the left regular action. With that convention, the generatorwise flatness condition is
\(\varphi_{0,j}\partial^F=\partial^F\varphi_{1,j},\, j=1,2.\)

To ensure reproducibility, we provide detailed computations of the code parameters $[[n,k,d]]$ for both the untwisted code and the code with all singular twists from Case 3. The details are given in Appendix C.

Case 1: Both twists are invertible. We choose twists

\[
\varphi_{0,1}=
\begin{bmatrix}
1&r+r^2\\
0&r
\end{bmatrix},
\qquad
\varphi_{1,1}=
\begin{bmatrix}
1&0\\
s+rs&r
\end{bmatrix},
\qquad
\varphi_{0,2}=
\varphi_{1,2}=
\begin{bmatrix}
0&r\\
s&0
\end{bmatrix}.
\]

A direct computation shows that the flatness condition holds: \(\varphi_{0,j}\partial^F=\partial^F\varphi_{1,j}.\)
Therefore \(\partial_1\partial_2=0\). So the twisted object is a valid chain complex.

Parameters of the twisted code

The qubits live on \(C_1=(B_1\otimes F_0)\oplus(B_0\otimes F_1).\)
Its \(R\)-rank is \(2\cdot 2+2\cdot 2=8.\)
Since \(\dim_{\mathbb F_2}R=|D_3|=6\), the binary blocklength is \(n=8\cdot 6=48\). So \(n=48\).

After expanding the total differentials to binary matrices via the regular representation, we get \(\operatorname{rank}(\partial_2)=21\) and \(\operatorname{rank}(\partial_1)=21.\)
Hence \(k=n-\operatorname{rank}(\partial_1)-\operatorname{rank}(\partial_2)=48-21-21=6\). So \(k=6\).

For the CSS distances: \(d_X=\min\{\mathrm{wt}(x):x\in \ker\partial_1\setminus \operatorname{im}\partial_2\}\),
\(d_Z=\min\{\mathrm{wt}(z):z\in \ker\partial_2^\top\setminus \operatorname{im}\partial_1^\top\}\). we find \(d_X=2\) and \(d_Z=2\), so \(d=2\).

Therefore the twisted code has parameters \([[48,6,2]]\).

Parameters of the corresponding untwisted lifted product code

For the construction of this code, we replace the twists by identities: \(\varphi_{0,1}=\varphi_{1,1}=\varphi_{0,2}=\varphi_{1,2}=I\). The same computation gives the parameters of the lifted product code \([[48,6,2]]\).

So this is an invertible-twist case where the twisted and untwisted constructions have the same \([[n,k,d]]\).

Next, we present two modified twisted code cases, each of which achieves a higher dimension \(k\) than the corresponding untwisted lifted product code while preserving the same distance \(d\).

Case 2: One twist is invertible, the other is not: We replace the second twist by \(\partial^F\). With our modified second twist
\[\varphi_{0,2}=\varphi_{1,2}=
\begin{bmatrix}1&r\\ s&1\end{bmatrix}
=\partial^F,\]

it is straightforward to verify that this twist is not invertible and therefore not an automorphism of \(R^2\). Nevertheless, the twisted complex still satisfies the chain condition, making it a valid complex. We now determine the parameters of this twisted code. Since \(C_1\) remains unchanged, \(n=48\). For the twisted complex, the induced binary boundary maps have ranks as follows: \(\operatorname{rank}(\partial_2)=19\) and \(\operatorname{rank}(\partial_1)=19\). Hence \(k=48-19-19=10.\)

The CSS distances are \(d_X=2\) and \(d_Z=2\), so \(d=2\). Thus the twisted code has parameters \([[48,10,2]]\). It has a better \(k\) than the untwisted lifted product code while keeping the same distance.


Case 3: Neither twist is invertible. We make both twists equal to \(\partial^F\) with
\[\partial^F=
\begin{bmatrix}1&r\\ s&1\end{bmatrix}\, \text{and}\,\varphi_{0,1}=\varphi_{1,1}=\varphi_{0,2}=\varphi_{1,2}=\partial^F.
\]
The twisted object is clearly a valid chain complex. Since chain \(C_1\) is unchanged, we have \(n=48\). Expanding the twisted boundary maps to binary matrices via the regular representation yields:
\(\operatorname{rank}(\partial_1)=18\) and \(\operatorname{rank}(\partial_2)=18.\)
Therefore \(k=48-18-18=12\). For the distances, we check the first nontrivial weights directly and find
\(d_X=2\) and \(d_Z=2\). Hence \(d=2\).
As a result, the twisted code achieves the parameters \([[48,12,2]]\), by contrast, the parameters for the corresponding untwisted lifted product code are  \([[48,6,2]].\)

Remark\, \, In this case, making both twist pairs equal to the noninvertible map \(\partial^F\) doubles the encoded dimension from \(6\) to \(12,\), while keeping the same length and the same minimum distance: \([[48,6,2]] \to [[48,12,2]].\)
So this is a clean finite-length example where the twisted construction is strictly better than the untwisted lifted product code in \(k\) at fixed \((n,d)\). 

Overall, this example suggests that noninvertible twists can act as a homology amplifier: singular twists may introduce additional logical degrees of freedom, and multiple such twists can contribute additively to the encoded dimension. The minimum distance, however, does not necessarily improve and may remain small. Thus, at least at the finite-length level, the twisted construction can increase the code rate without decreasing the distance in the example considered here.

\section{Concluding Remarks}\label{conclusion}
In this paper, we introduced a twisted fiber bundle construction of quantum CSS codes over the group algebra \(R=\mathbb{F}_2[G]\) for arbitrary finite groups $G$. The construction assigns to each base generator a generator-dependent $R$-linear twist satisfying a flatness condition, ensuring that the resulting total complex forms a chain complex. The usual lifted product construction is recovered as the untwisted special case in which all twists are identities.

We proved that when the fiber twists are invertible, the twisted complex is chain-isomorphic to the corresponding untwisted complex. It follows that the associated binary CSS codes have the same blocklength $n$ and encoded dimension $k$, and in many cases are equivalent up to a coordinate transformation. By contrast, allowing the twists to be merely chain-compatible endomorphisms can reduce the rank of the induced boundary maps and thereby generate additional logical degrees of freedom.

Explicit examples over \(R=\mathbb{F}_2[D_3]\) show that noninvertible, chain-compatible twists can yield twisted codes with strictly larger $k$ than their untwisted lifted product counterparts, while preserving the blocklength $n$ and, in the examples considered here, the minimum distance $d$. These constructions provide concrete finite-length evidence that twisted fiber bundle codes with generator-dependent $R$-module twists are genuinely more flexible than lifted product codes and that singular twisting can substantially enlarge the CSS code design space.

An important open problem is whether invertibility together with chain-compatibility is sufficient to force equality of the full CSS parameters 
$[[n,k,d]]$. Our results settle this question for $n$ and $k$, but the distance $d$ remains unresolved in general. Addressing this issue will require either proving that the induced middle-space isomorphism preserves minimum logical weight or identifying a counterexample in which the distance changes.

\section{Acknowledgments}

In the preparation of this work, the author made use of GPT-5.3 Instant and GPT-5.4 Thinking for grammar and clarity refinement, as well as to aid in computing the code parameters presented in the examples. These interactions helped improve the quality of this paper. After using this tool, the author carefully reviewed and edited the content as necessary and takes full responsibility for the final content of this paper.



\section{Appendices}

\begin{center}
{\bf Appendix  A} Proof of the chain condition $H_XH_Z^{\mathsf T}=0$ 
\end{center}

We justify the CSS chain condition after passing from the formal \(R\)-module sequence to its binary expansion via compatible left and right regular representations.

Recall that the formal boundary maps are given by
\[
\partial_1=
\begin{bmatrix}
\partial^B_{\varphi_0} & \mathsf{id}_{B_0}\otimes \partial^F
\end{bmatrix},
\qquad
\partial_2=
\begin{bmatrix}
\mathsf{id}_{B_1}\otimes \partial^F \\
\partial^B_{\varphi_1}
\end{bmatrix}.
\]
Thus, symbolically, \(\partial_1\partial_2=\partial^B_{\varphi_0}(\mathsf{id}_{B_1}\otimes \partial^F)+(\mathsf{id}_{B_0}\otimes \partial^F) \partial^B_{\varphi_1}.\)

At the purely symbolic \(R\)-module level, this product need not vanish when \(R\) is noncommutative. Indeed, if \(b_{ij}:=(\partial^B)_{ij}\), then the corresponding local contribution has the form \(b_{ij}\,\varphi_{0,j}\partial^F+\partial^F\, b_{ij}\varphi_{1,j}.\)
Using the flatness condition \(\varphi_{0,j}\partial^F=\partial^F\varphi_{1,j},\)
this becomes
\[
b_{ij}\,\partial^F\varphi_{1,j}
+
\partial^F\, b_{ij}\varphi_{1,j}.
\]
In general, this expression is not guaranteed to vanish over a noncommutative algebra \(R\), since \(b_{ij}\) need not commute with \(\partial^F\).

We now show that the desired chain condition does hold after binary expansion. Fix a basis of \(R\) over \(\mathbb F_2\). For \(r\in R\), let \(\rho_r:x\mapsto xr\) and \(\lambda_r:x\mapsto rx\)
denote the right and left regular representations, respectively. These two actions commute:
\[
\rho_{r_1}\lambda_{r_2}=\lambda_{r_2}\rho_{r_1}
\qquad
\text{for all } r_1, r_2\in R.
\]
In the binary expansion used in this construction, entries coming from the base boundary map \(\partial^B\) are replaced by right regular representation matrices, whereas entries coming from the fiber boundary map \(\partial^F\) and from the twists \(\varphi_{0/1,j}\) are replaced by left regular representation matrices.

Let \(\widehat{\partial^F}\), \(\widehat{\varphi_{0,j}}\), and \(\widehat{\varphi_{1,j}}\) denote the binary matrices obtained from \(\partial^F\), \(\varphi_{0,j}\), and \(\varphi_{1,j}\), respectively, by left regular expansion. The flatness condition gives \(\widehat{\varphi_{0,j}}\widehat{\partial^F}=\widehat{\partial^F}\widehat{\varphi_{1,j}}.\)

The binary expansion of the local contribution above is therefore
\[
\rho_{b_{ij}}\widehat{\varphi_{0,j}}\widehat{\partial^F}
+
\widehat{\partial^F}\rho_{b_{ij}}\widehat{\varphi_{1,j}}.
\]
Using the expanded flatness condition and the commutation of the left and right regular representations, we obtain
\[
\begin{aligned}
\rho_{b_{ij}}\widehat{\varphi_{0,j}}\widehat{\partial^F}
+
\widehat{\partial^F}\rho_{b_{ij}}\widehat{\varphi_{1,j}}
&=
\rho_{b_{ij}}\widehat{\partial^F}\widehat{\varphi_{1,j}}
+
\widehat{\partial^F}\rho_{b_{ij}}\widehat{\varphi_{1,j}}  \\
&=
\rho_{b_{ij}}\widehat{\partial^F}\widehat{\varphi_{1,j}}
+
\rho_{b_{ij}}\widehat{\partial^F}\widehat{\varphi_{1,j}} \\
&=0,
\end{aligned}
\]
where the last equality holds because the computation is over \(\mathbb F_2\).

Hence every local block contribution vanishes after binary expansion. Therefore the expanded boundary matrices compose to zero. If \(H_X\) is the binary matrix obtained from \(\partial_1\) and \(H_Z\) is the binary matrix obtained from \(\partial_2^{\mathsf T}\), then
\[
H_XH_Z^{\mathsf T}=0.
\]
Thus the binary-expanded construction defines a valid CSS code.

\begin{center}
{\bf Appendix B} Proof of chain-isomorphism between two complexes
\end{center}
We define $R$-linear maps
\(T_2:C_2^{\mathrm{untw}}\to C_2^{\mathrm{tw}}\), \(T_1:C_1^{\mathrm{untw}}\to C_1^{\mathrm{tw}}\) and \(T_0:C_0^{\mathrm{untw}}\to C_0^{\mathrm{tw}}\)
by \(T_2(e_j\otimes x)=e_j\otimes \varphi_{1,j}^{-1}(x)\), \(T_1(e_j\otimes y,\;u)=\bigl(e_j\otimes \varphi_{0,j}^{-1}(y),\;u\bigr)\) and \(T_0=\mathrm{id}_{B_0\otimes_R F_0}\).  Since each $\varphi_{0/1,j}$ is invertible, each $T_i$ is an isomorphism.

We first verify \(\partial_2^{\mathrm{tw}}T_2=T_1\partial_2^{\mathrm{untw}}\). For a basis tensor $e_j\otimes x\in B_1\otimes_R F_1$,
\(T_2(e_j\otimes x)=e_j\otimes \varphi_{1,j}^{-1}(x),\)
hence \(\partial_2^{\mathrm{tw}}T_2(e_j\otimes x)=\left(e_j\otimes \partial^F\varphi_{1,j}^{-1}(x),\;\sum_i (\partial^B)_{ij}e_i'\otimes x\right).\)
From \(\varphi_{0,j}\partial^F=\partial^F\varphi_{1,j}\) and invertibility, we obtain \(\partial^F\varphi_{1,j}^{-1}=\varphi_{0,j}^{-1}\partial^F.\)
Therefore
\[
\partial_2^{\mathrm{tw}}T_2(e_j\otimes x)
=
\left(
e_j\otimes \varphi_{0,j}^{-1}\partial^F(x),\;
\sum_i (\partial^B)_{ij}e_i'\otimes x
\right).
\]
On the other hand, \(\partial_2^{\mathrm{untw}}(e_j\otimes x)=\left(e_j\otimes \partial^F(x),\;\sum_i (\partial^B)_{ij}e_i'\otimes x\right),\)
and then \(T_1\partial_2^{\mathrm{untw}}(e_j\otimes x)=\left(e_j\otimes \varphi_{0,j}^{-1}\partial^F(x),\;\sum_i (\partial^B)_{ij}e_i'\otimes x\right).\)
Thus \(\partial_2^{\mathrm{tw}}T_2=T_1\partial_2^{\mathrm{untw}}.\)

Next we verify \(\partial_1^{\mathrm{tw}}T_1=T_0\partial_1^{\mathrm{untw}}.\)
For $e_j\otimes y\in B_1\otimes_R F_0$, \(T_1(e_j\otimes y)=e_j\otimes \varphi_{0,j}^{-1}(y),\)
so \(\partial_1^{\mathrm{tw}}T_1(e_j\otimes y)=\sum_i (\partial^B)_{ij}e_i'\otimes y=(\partial^B\otimes \mathrm{id}_{F_0})(e_j\otimes y)=T_0\partial_1^{\mathrm{untw}}(e_j\otimes y).\)
For $u\in B_0\otimes_R F_1$, the map $T_1$ is the identity on the second summand, hence
\[
\partial_1^{\mathrm{tw}}T_1(u)
=
(\mathrm{id}_{B_0}\otimes \partial^F)(u)
=
T_0\partial_1^{\mathrm{untw}}(u).
\]
Therefore $(T_2,T_1,T_0)$ is a chain isomorphism. In particular, \(H_1(C^{\mathrm{tw}}_\bullet)\cong H_1(C^{\mathrm{untw}}_\bullet).\)


\begin{center}
{\bf Appendix C}. Computation of Code Parameters for the Untwisted Code and the Code with All Singular Twists from Case 3
\end{center}

For the group algebra \(\mathbb F_2[D_3]\), we choose the ordered basis \(\{1,r,r^2,s,sr,sr^2\}\) for \(D_3\). With respect to this basis, the right and left regular representation matrices needed in the computations are as follows:
\[
\rho_r=
\begin{bmatrix}
0&0&1&0&0&0\\
1&0&0&0&0&0\\
0&1&0&0&0&0\\
0&0&0&0&0&1\\
0&0&0&1&0&0\\
0&0&0&0&1&0
\end{bmatrix},
\qquad
\lambda_r=
\begin{bmatrix}
0&0&1&0&0&0\\
1&0&0&0&0&0\\
0&1&0&0&0&0\\
0&0&0&0&1&0\\
0&0&0&0&0&1\\
0&0&0&1&0&0
\end{bmatrix},
\qquad
\lambda_s=
\begin{bmatrix}
0&0&0&1&0&0\\
0&0&0&0&1&0\\
0&0&0&0&0&1\\
1&0&0&0&0&0\\
0&1&0&0&0&0\\
0&0&1&0&0&0
\end{bmatrix}.
\]
Moreover,
\[
\rho_{r+r^2}=
\begin{bmatrix}
0&1&1&0&0&0\\
1&0&1&0&0&0\\
1&1&0&0&0&0\\
0&0&0&0&1&1\\
0&0&0&1&0&1\\
0&0&0&1&1&0
\end{bmatrix},
\qquad
\rho_{1+r+r^2}=
\begin{bmatrix}
1&1&1&0&0&0\\
1&1&1&0&0&0\\
1&1&1&0&0&0\\
0&0&0&1&1&1\\
0&0&0&1&1&1\\
0&0&0&1&1&1
\end{bmatrix},
\qquad
\lambda_1=
\begin{bmatrix}
1&0&0&0&0&0\\
0&1&0&0&0&0\\
0&0&1&0&0&0\\
0&0&0&1&0&0\\
0&0&0&0&1&0\\
0&0&0&0&0&1
\end{bmatrix}.
\]

\subsection*{C.1. Computation of the Code Parameters \([[n,k,d]]\) for the Untwisted Code}

For the untwisted code, the binary-expanded boundary maps are obtained from
\[
\partial_1=
\begin{bmatrix}
\partial^B\otimes \mathrm{id}_{F_0} & \mathrm{id}_{B_0}\otimes \partial^F
\end{bmatrix}
\text{and}\,\,
\partial_2=
\begin{bmatrix}
\mathrm{id}_{B_1}\otimes \partial^F\\
\partial^B\otimes \mathrm{id}_{F_1}
\end{bmatrix}.
\]
Explicitly, after applying the appropriate right and left regular representations, we obtain
\[
\partial_1=
\begin{bmatrix}
0&0&\rho_{r+r^2}&0&\lambda_1&\lambda_r&0&0\\
0&0&0&\rho_{r+r^2}&\lambda_s&\lambda_1&0&0\\
\rho_{1+r+r^2}&0&0&0&0&0&\lambda_1&\lambda_r\\
0&\rho_{1+r+r^2}&0&0&0&0&\lambda_s&\lambda_1
\end{bmatrix}
\]
and
\[
\partial_2=
\begin{bmatrix}
\lambda_1&\lambda_r&0&0\\
\lambda_s&\lambda_1&0&0\\
0&0&\lambda_1&\lambda_r\\
0&0&\lambda_s&\lambda_1\\
0&0&\rho_{r+r^2}&0\\
0&0&0&\rho_{r+r^2}\\
\rho_{1+r+r^2}&0&0&0\\
0&\rho_{1+r+r^2}&0&0
\end{bmatrix}.
\]

From the block structures of \(\partial_1\) and \(\partial_2\), we have
\(
\operatorname{rank}(\partial_1)=\operatorname{rank}(\partial_2).
\)
Moreover,
\(
\operatorname{rank}(\partial_1)
=
\operatorname{rank}(W_1)+\operatorname{rank}(W_2),
\)
where
\[
W_1=
\begin{bmatrix}
\rho_{r+r^2}&0&\lambda_1&\lambda_r\\
0&\rho_{r+r^2}&\lambda_s&\lambda_1
\end{bmatrix},
\qquad
W_2=
\begin{bmatrix}
\rho_{1+r+r^2}&0&\lambda_1&\lambda_r\\
0&\rho_{1+r+r^2}&\lambda_s&\lambda_1
\end{bmatrix}.
\]
A direct computation over \(\mathbb F_2\) gives
\(
\operatorname{rank}(W_1)=11\) and 
\(\operatorname{rank}(W_2)=10.
\)
Hence
\(
\operatorname{rank}(\partial_1)
=
\operatorname{rank}(\partial_2)
=
11+10=21.
\)
Since the length is \(n=48\), the number of encoded qubits is
\[
k
=
n-\operatorname{rank}(\partial_1)-\operatorname{rank}(\partial_2)
=
48-2(21)
=
6.
\]

We next compute the minimum distance. Recall that
\(
d=\min(d_X,d_Z),
\)
where
\(
d_X
=
\min\left\{
\operatorname{wt}(x):
x\in \ker(\partial_1)\setminus \operatorname{im}(\partial_2)
\right\}
\)
and
\(
d_Z
=
\min\left\{
\operatorname{wt}(z):
z\in \ker(\partial_2^{\mathsf T})\setminus \operatorname{im}(\partial_1^{\mathsf T})
\right\}.
\)

From the explicit forms of \(\partial_1\) and \(\partial_2\), every column of \(\partial_1\) and every row of \(\partial_2\) contains at least one nonzero binary entry. Therefore no vector of weight one lies in \(\ker(\partial_1)\), and no vector of weight one lies in \(\ker(\partial_2^{\mathsf T})\). It follows that
\(
d_X\ge 2\)
and
\(d_Z\ge 2.
\)

On the other hand, direct verification shows that the weight-two vector
\(
x=(1,1,0,\ldots,0)
\)
satisfies
\(
x\in \ker(\partial_1)\setminus \operatorname{im}(\partial_2),
\)
and the weight-two vector
\(
z=(0,\ldots,0,1,1)
\)
satisfies
\(
z\in \ker(\partial_2^{\mathsf T})\setminus \operatorname{im}(\partial_1^{\mathsf T}).
\)
Thus
\(
d_X\le 2\) and \(d_Z\le 2\).
Consequently,
\(
d_X=d_Z=2\) and \(d=2\). 

Therefore the untwisted code has parameters \([[48,6,2]].\)

\subsection*{C.2. Computation of the Code Parameters \([[n,k,d]]\) for the Code with All Singular Twists from Case 3}

We recall the input data for the twisted code from Case 3:
\(
\partial^B=
\begin{bmatrix}
0 & r+r^2\\
1+r+r^2 & 0
\end{bmatrix}\,\, \text{and}\,\,
\partial^F=\varphi_1=\varphi_2=
\begin{bmatrix}
1&r\\
s&1
\end{bmatrix}.
\)

The corresponding twisted boundary maps are
\(
\partial_1=
\begin{bmatrix}
\partial^B_{\varphi_0} & \mathrm{id}_{B_0}\otimes \partial^F
\end{bmatrix}
\text{and}\,\,
\partial_2=
\begin{bmatrix}
\mathrm{id}_{B_1}\otimes \partial^F\\
\partial^B_{\varphi_1}
\end{bmatrix}.
\)

After binary expansion, these become
\[
\partial_1=
\begin{bmatrix}
0&0&\rho_{r+r^2}\lambda_1&\rho_{r+r^2}\lambda_r&\lambda_1&\lambda_r&0&0\\
0&0&\rho_{r+r^2}\lambda_s&\rho_{r+r^2}\lambda_1&\lambda_s&\lambda_1&0&0\\
\rho_{1+r+r^2}\lambda_1&\rho_{1+r+r^2}\lambda_r&0&0&0&0&\lambda_1&\lambda_r\\
\rho_{1+r+r^2}\lambda_s&\rho_{1+r+r^2}\lambda_1&0&0&0&0&\lambda_s&\lambda_1
\end{bmatrix}
\]
and
\[
\partial_2=
\begin{bmatrix}
\lambda_1&\lambda_r&0&0\\
\lambda_s&\lambda_1&0&0\\
0&0&\lambda_1&\lambda_r\\
0&0&\lambda_s&\lambda_1\\
0&0&\rho_{r+r^2}\lambda_1&\rho_{r+r^2}\lambda_r\\
0&0&\rho_{r+r^2}\lambda_s&\rho_{r+r^2}\lambda_1\\
\rho_{1+r+r^2}\lambda_1&\rho_{1+r+r^2}\lambda_r&0&0\\
\rho_{1+r+r^2}\lambda_s&\rho_{1+r+r^2}\lambda_1&0&0
\end{bmatrix}.
\]

As in the untwisted case, the block structures imply
\(
\operatorname{rank}(\partial_1)=\operatorname{rank}(\partial_2)
\)
and
\(
\operatorname{rank}(\partial_1)
=
\operatorname{rank}(V_1)+\operatorname{rank}(V_2),
\)
where
\(
V_1=
\begin{bmatrix}
\rho_{r+r^2}\lambda_1&\rho_{r+r^2}\lambda_r&\lambda_1&\lambda_r\\
\rho_{r+r^2}\lambda_s&\rho_{r+r^2}\lambda_1&\lambda_s&\lambda_1
\end{bmatrix}
\text{and}\,\,
V_2=
\begin{bmatrix}
\rho_{1+r+r^2}\lambda_1&\rho_{1+r+r^2}\lambda_r&\lambda_1&\lambda_r\\
\rho_{1+r+r^2}\lambda_s&\rho_{1+r+r^2}\lambda_1&\lambda_s&\lambda_1
\end{bmatrix}.
\)

A direct computation over \(\mathbb F_2\) gives
\(
\operatorname{rank}(V_1)=9\) and \(\operatorname{rank}(V_2)=9.\)
Therefore
\(
\operatorname{rank}(\partial_1)
=
\operatorname{rank}(\partial_2)
=
9+9=18.
\)

Since \(n=48\), we obtain
\(
k
=
n-\operatorname{rank}(\partial_1)-\operatorname{rank}(\partial_2)
=
48-2(18)
=
12.
\)

It remains to compute the distance. From the explicit forms of \(\partial_1\) and \(\partial_2\), every column of \(\partial_1\) and every row of \(\partial_2\) contains at least one nonzero binary entry. Hence no weight-one vector lies in \(\ker(\partial_1)\), and no weight-one vector lies in \(\ker(\partial_2^{\mathsf T})\). Thus
\(
d_X\ge 2\) and \(d_Z\ge 2.\)

On the other hand, direct verification shows that the weight-two vector
\(
x=(1,1,0,\ldots,0)
\)
satisfies
\(
x\in \ker(\partial_1)\setminus \operatorname{im}(\partial_2),
\)
and the weight-two vector
\(
z=(0,\ldots,0,1,1)
\)
satisfies
\(
z\in \ker(\partial_2^{\mathsf T})\setminus \operatorname{im}(\partial_1^{\mathsf T}).
\)
Therefore
\(
d_X\le 2\) and \(d_Z\le 2\).
Combining the lower and upper bounds gives
\(
d_X=d_Z=2\) and \(d=2\). 

Thus the twisted code with all singular twists from Case 3 has parameters
\(
[[48,12,2]].
\)

Finally, we note that the code parameters in the remaining cases can be verified by the same procedure.

\begin{center}
{\bf Appendix D} Lifted product codes construction (Panteleev–Kalachev \cite{PK2019, PK2021})
\end{center}

A lifted product code, denoted by \(\mathrm{LP}(A,B)\), is constructed from two classical linear codes \(C_A\) and \(C_B\), represented by their parity-check matrices \(A\) and \(B\). These matrices are interpreted as boundary maps of 2-term chain complexes of free modules over an associative \(\mathbb{F}_2\)-algebra \(R\) with identity, typically a group algebra \(R = \mathbb{F}_2[G]\) for a finite group \(G\). In practice, such 2-term chain complexes over \(R\) often arise from 2-term chain complexes of \(\mathbb{F}_2\)-vector spaces equipped with a free action of \(G\).

Below is a step-by-step outline of the standard construction of a lifted product code using matrices over a group algebra \(R = \mathbb{F}_2[G]\).

1. Choose the base parameters

Select a finite group \(G\) with \(|G|=\ell\) (e.g., the cyclic group \(\mathbb{Z}_\ell)\) and form the group algebra
\(R = \mathbb{F}_2[G].\)

2. Define the base matrices over \(R\)

We choose matrices
\(
A \in \mathcal{M}_{m_A \times n_A}(R)\) and \(B \in \mathcal{M}_{m_B \times n_B}(R)\) whose entries lie in \(R\). These matrices serve as boundary maps of 2-term chain complexes \(\mathcal{A}: R^{n_A}\xrightarrow {A} R^{m_A}\) and \(\mathcal{B}:R^{n_B}\xrightarrow{B} R^{m_B}\) over the group algebra \(R = \mathbb{F}_2[G]\), and they may be viewed as “lifted” versions of classical parity-check matrices or protographs.

3. Form the hypergraph-product–style matrices over \(R\)

We take the standard hypergraph product, using Kronecker products over \(R\), to obtain
\[H_X =\begin{bmatrix}A \otimes I_{m_B} & I_{m_A} \otimes B\end{bmatrix}\,\text{and}\,\, H_Z =
\begin{bmatrix} I_{n_A} \otimes B^{\mathsf T} & A^{\mathsf T} \otimes I_{n_B}\end{bmatrix},\]
where \(\otimes\) denotes the Kronecker product over \(R\), and \(I_k\) denotes the \(k \times k\) identity matrix with entries in \(R\).
The matrices \(H_X\) and \(H_Z^{\mathsf T}\) represent the boundary maps of the lifted product \(\mathcal{A} \otimes_R \mathcal{B}\) of the two-term chain complexes \(\mathcal{A}\) and \(\mathcal{B}\). Explicitly, this yields the chain complex
\(R^{n_A n_B}\xrightarrow{H_Z^{\mathsf T}}R^{n_A m_B} \oplus R^{m_A n_B}\xrightarrow{H_X}R^{m_A m_B}.\)
It is crucial that all four Kronecker products are computed over \(R\) at the matrix level before any reduction of entries to \(\mathbb{F}_2\) (sometimes termed binary expansion over \(\mathbb{F}_2\)) is performed.

After forming these products, we write
\(
H_X = (r^{x}_{ij})\) and \(H_Z = (r^{z}_{ij})\) where each nonzero \(r^x_{ij}\) or \( r^z_{ij}\) is either an entry of \(A\) or an entry of \(B\).

4. Reduce the \(R\)-matrices back to \(\mathbb{F}_2\) to form the lifted product code parity-check matrices

To obtain binary parity-check matrices, we reduce the matrices \(H_X\) and \(H_Z\) from \(R\) back to \(\mathbb{F}_2\). For each entry \(r \in R\), we replace it by its corresponding matrix representation: the right regular representation for an entry from \(A\); the left regular representation for an entry from \(B\); zero matrix for entry 0.

This yields block matrices
\(
\hat{H}_X = (\hat{r}^x_{ij})\,\, \text{and}\,\,
\hat{H}_Z = (\hat{r}^z_{ij})
\)
where each \(\hat{r}\) is an \(\ell \times \ell\) binary matrix.

One checks that these matrices satisfy the CSS orthogonality condition
\(
\hat{H}_X\hat{H}_Z^{,T} = 0.
\)
Thus the binary matrices \(\hat{H}_X\) and \(\hat{H}_Z\) define a CSS quantum code, called the lifted product code, denoted \(\mathrm{LP}(A,B)\).

\begin{center}
{\bf Appendix E: Fiber bundle codes construction (Hastings–Haah–O’Donnell \cite{HHD2021})}
\end{center}
A fiber bundle code is a CSS quantum LDPC code constructed by attaching a fixed local “fiber” code to each vertex of a base complex and coupling these fibers along base edges via a discrete group action. Algebraically, it is defined by a twisted tensor-product (or total) chain complex, where the twisting encodes parallel transport of the fiber along the base. This structure generalizes hypergraph product codes and enables improved distance through nontrivial bundle holonomy.

Below is a standard, step-by-step outline of the fiber bundle code construction starting from a 2-term base complex \(B:\, B_1 \xrightarrow{\partial^B} B_0\) and a 2-term fiber complex \(F:\, F_1 \xrightarrow{\partial^F} F_0,\) following the formulation used in Hastings–Haah–O’Donnell \cite{HHD2021} and subsequent refinements, e.g. \cite{BE2021}

We will emphasize what data are chosen, how the total complex is formed, and how the CSS code is read off.

1. Input data and assumptions

 (a) Base complex  \(B:\, B_1 \xrightarrow{\partial^B} B_0\) where \(B_0\) and \(B_1\) are finite-dimensional \(\mathbb{F}_2\)-vector spaces. We think of \(B_0\) as base vertices and \(B_1\) as base edges.

(b) Fiber complex \(F:\, F_1 \xrightarrow{\partial^F} F_0\) where \(F_0\) and \( F_1\) are finite-dimensional \(\mathbb{F}_2\)-vector spaces. This defines the local code attached to each base vertex.

(c) Structure group action (twisting data)  Choose a group \(G\) (typically finite) and a representation \(\rho_i: G \to \mathrm{Aut}(F_i)\) where \(i \in \{0,1\}\) by chain automorphisms: \( \rho_0(g)\partial^F = \partial^F \rho_1(g).\)

(d) Bundle connection on the base \quad For each oriented base edge \(b^1 \in B_1\) and its base vertex \(b^0\in \partial^Bb^1\), choose \(g_{(b^1, b^0)} \in G,\) interpreted as parallel transport along \(b^1\) at \(b^0\). Accordingly, we define connections/twists $\varphi^i(b^1,b^0)\triangleq\rho_i(g_{(b^1,b^0)})\in \mathrm{Aut}(F_i)$ that commutes with $\partial^F$ (i.e., $\varphi^0(b^1,b^0)\circ \partial^F=\partial^F\circ \varphi^1(b^1,b^0))$. The differential $\partial^{F_i}_{\varphi}: B_1\otimes F_i \rightarrow B_0\otimes F_i$ is given: \(\partial^{F_i}_{\varphi}(b^1\otimes f^i)=\sum_{b^0\in \partial^B b^1}b^0\otimes \varphi^i(b^1,b^0)(f^i)\) where $f^i$ is the basis vectors of $F_i$.



2. Form the graded total space (qubits and checks)

Define the total chain groups:\[
\begin{aligned}
C_2: &= B_1 \otimes F_1, \\
C_1: &= (B_1 \otimes F_0)\oplus(B_0 \otimes F_1), \\
C_0: &= B_0 \otimes F_0
\end{aligned}
\]


where qubits live in \(C_1\), \(Z\)-checks come from \(C_2\), and \(X\)-checks come from \(C_0\).

3.. Define the twisted boundary map \(\partial_2: C_2 \to C_1\) 
where \(\partial_2=\begin{bmatrix}
\mathrm{id}_{B_1}\otimes \partial^F\\
 \partial^{F_1}_{\varphi}
\end{bmatrix}\)


The upper block represents the fiber boundary taken over a fixed base edge. The lower block encodes the base boundary together with the transported fiber data, where the transport is determined by the twisting (or monodromy) \(\varphi\). This term captures precisely how the bundle structure enters the construction.

4. Define the twisted boundary map \(\partial_1: C_1 \to C_0\)
 with \(\partial_1=\begin{bmatrix}\partial^{F_0}_{\varphi}, \mathrm{id}_{B_0}\otimes \partial^F \end{bmatrix}\)
 
The right block represents the fiber boundary taken over a fixed base vertex. The left block encodes the base boundary together with the transported fiber data, where the transport is determined by the twisting (or monodromy) \(\varphi\). This term captures precisely how the bundle structure enters the construction.







5. Verify chain complex condition

One can check \(\partial_1 \circ \partial_2 = 0\) by using \((\partial^B)^2 = 0\), \((\partial^F)^2 = 0\), and  \(\rho(g)\) commuting with \(\partial^F\).

Hence \(C_2 \xrightarrow{\partial_2} C_1 \xrightarrow{\partial_1} C_0\) is a valid chain complex, denoted by \(B\otimes_{\varphi}F\).

6. Extract the CSS code

The fiber bundle CSS code is defined by parity check matrices $H_X$ and $H_Z$ where $H_X$ and $H_Z$ are the respective matrix representations of $\partial_1$ and $\partial_2^\top$.


Parameters: \(n=\dim C_1, k=\dim H_1(C), \,\text{and}\,\, d=\mathrm{min}(d_X, d_Z)\) where $d_X, d_Z$ are determined by minimal nontrivial cycles/cocycles.

Remark: If \(g_{(b^1,b^0)} = 1\) for all pairs \((b^1,b^0)\), \(C \cong B \otimes F\) becomes an ordinary hypergraph product complex, the code is the hypergraph product code. Otherwise, a nontrivial bundle may produce longer logical operators and improved distance for the fiber bundle code.

This construction can be depicted in Figure 1.

\begin{tikzpicture}
    \node (A) at (0,0) {$B_1\otimes F_1$};
    \node (B) at (0,-3) {$B_1\otimes F_0$};
    \node (C) at (7,0) {$B_0\otimes F_1$};
    \node (D) at (7,-3) {$B_0\otimes F_0$};

    \draw[->] (A) -- (B) node[midway,left] {$\mathrm{id}_{B_1}\otimes \partial^F$};
    \draw[->] (C) -- (D) node[midway,right] {$\mathrm{id}_{B_0}\otimes \partial^F$};
     \draw[->] (A) -- (C) node[midway,above] {$\partial^{F_1}_{\varphi}$};
    \draw[->] (B) -- (D) node[midway,below] {$\partial^{F_0}_{\varphi}$};

    \node[text width=8cm] at (3.5,-4) {Figure 1:The fiber bundle double complex $B\otimes_{\varphi} F$};
\end{tikzpicture}

\end{footnotesize}

\end{document}